\documentclass[twocolumn,pre,amsmath,amssymb]{revtex4}
 
\usepackage{graphicx}
\usepackage{dcolumn}
\usepackage{bm}
 
\newcommand{\bfM}{{\mathbf{M}}}

\newcommand{\bfN}{{\mathbf{N}}}
\newcommand{\dotbfM}{\dot{\mathbf{M}}}

\newcommand{\dotbfN}{\dot{\mathbf{N}}}
\newcommand{\dotrho}{\dot{\rho}}
\newcommand{\bfB}{{\mathbf{B}}}

\newcommand{\atan}{{\mathrm{arctan}}}
\newcommand{\dd}{{\mathrm{d}}}
\newcommand{\ee}{{\mathrm{e}}}
\newcommand{\bargamma}{{\bar{\gamma}}}
\newcommand{\barx}{{\bar{x}}}
 
\begin{document}
 
\preprint{APS/123-QED}
 
\title{Analytical solutions for two-level systems with damping}
 
\author{Christian Brouder}
\affiliation{%
Institut de Min\'eralogie et de Physique des Milieux Condens\'es,
CNRS UMR7590,\\
Universit\'es Paris 6 et 7, IPGP, 140 rue de Lourmel,
F-75015 Paris, France.
}%
 
\date{\today}
 
\begin{abstract}
A method is proposed to transform any analytic solution of
the Bloch equation into an analytic solution of the
Landau-Lifshitz-Gilbert equation. This allows for the
analytical description of the dynamics of a two level
system with damping.  This method shows that damping turns
the linear Schr\"odinger equation of
a two-level system into a nonlinear Schr\"odinger equation.
As applications, it is shown that damping has a relatively mild
influence on self-induced transparency but 
destroys dynamical localization.
\end{abstract}
 
\maketitle

Two-level systems have become almost
ubiquitous in modern physics. They are found for instance
in laser physics,
magnetic resonance spectroscopies, quantum computers, 
quantum teleportation and optoelectronics.
The state of a two-level system can be described by
an effective moment $\bfM$ and its dynamics by the 
equation $\dotbfM=-\gamma \bfM\times\bfB$, where $\gamma$
is the gyromagnetic factor and $\bfB$ is
an external time-dependent field.
We follow the standard (and inappropriate) custom of calling 
$\dotbfM=-\gamma \bfM\times\bfB$ the ``Bloch equation''.

The main drawback of the Bloch equation is its absence of
damping. There are many phenomenological models
of damping for two-level systems. When decoherence is
small, the system remains in a pure state, the length of 
$\bfM$ is constant and
damping is taken into account by the so-called 
Landau-Lifshitz-Gilbert (LLG) 
equation\cite{LandauLifshitz,Gilbert,Gilbert04}
\begin{eqnarray}
\dotbfM &=& -\gamma \bfM\times\bfB +\frac{\alpha}{M} \bfM\times\dotbfM,
\label{LLGeq}
\end{eqnarray}
where $M=|\bfM|$, $\dotbfM=\dd\bfM/\dd t$ and $\alpha > 0$.
It was shown that the LLG equation provides a realistic model of 
ferromagnetic resonance, micromagnetics,
spin-valve dynamics \cite{Emley}, the magnetism of thin films
\cite{HePB} and nanomagnets \cite{Ortigoza}, the dynamics of
domain walls in various geometries \cite{Barnes}.
Cheng and coll. recently proved that the LLG equations could
be derived from a microscopic model \cite{ChengJalil}.

The linear nature of the Bloch equation allowed for the
discovery of many analytic solutions. For example, a 
recent work identified 26 families of solutions that can
be expressed in terms of special functions \cite{Bagrov}
and significant steps toward the general solution were
carried out \cite{Kobayashi}.
Even when analytic solutions are not available, powerful
analytic approximation methods exist
\cite{Autler,Bialynicki,BarataPRL}.

By contrast, very few solutions of the physically more accurate
LLG equation are known.  In this paper, we describe a method by 
which \emph{any} analytic
solution of the Bloch equation can be transformed into
an analytic solution of the LLG equation.
Similarly, any analytic approximate solution of the Bloch equation
is transformed into an analytic approximate solution of
the LLG equation.  Then, we show that this transformation
turns the linear Schr\"odinger equation for a two-level
system into a non linear Schr\"odinger equation. Finally,
we investigate the influence of damping on 
self-induced transparency and dynamical localization.

We describe now the transformation from the solution
of the Bloch equation to the solution of the 
LLG equation.
Consider a solution of the Bloch equation
$\dotbfM(\gamma)=-\gamma \bfM(\gamma)\times \bfB$,
where the dependence of $M$ on the gyromagnetic 
factor was written explicitly and where
$\bfB$ is a real function of $t$.
Assume now that $\bfM(\gamma)$ is an
\emph{analytic function} of $\gamma$.
This allows us to define
$\bfN=\bfM(\bargamma)$, where $\bargamma=\gamma/(1-i\alpha)$.
This gives us
\begin{eqnarray*}
\dotbfN &=& \dotbfM(\bargamma)
=-\bargamma  \bfM(\bargamma)\times \bfB
=-\frac{\gamma}{1-i\alpha}  \bfN\times \bfB.
\end{eqnarray*}
The equation of motion implies that $M^2=\sum_i M_i^2$ does
not depend on $t$.
We define now
\begin{eqnarray}
\xi &=& \frac{N_x+iN_y}{M+N_z},\label{defxin}
\end{eqnarray}
If we calculate the derivative of $\xi$  with respect to $t$,
taking account of the fact that $M$ does not depend on time,
we find
\begin{eqnarray}
\dot{\xi} &=& \frac{\dot{N}_x+i\dot{N}_y}{M+N_z}-
     \frac{(N_x+iN_y)\dot{N}_z}{(M+N_z)^2}
\label{dotxiN}
\end{eqnarray}
If we substitute the equation of motion for $\dotbfN$,
we can check that $\xi$ satisfies
\begin{eqnarray}
\dot{\xi} &=& 
\frac{i \gamma}{2(1-i\alpha)}
\big(B^-\xi^2 +2 B_z \xi -B^+\big),
\label{dotxiN2}
\end{eqnarray}
where $B^\pm=B_x\pm i B_y$.
The complex function $\xi$ is used to define a
real vector $\bfM'$ by
\begin{eqnarray*}
M'_x &=& \frac{\xi+\xi^*}{|\xi|^2+1}M,\\
M'_y &=& -i\frac{\xi-\xi^*}{|\xi|^2+1}M,\\
M'_z &=& \frac{1-|\xi|^2}{|\xi|^2+1}M,
\end{eqnarray*}
so that we still have
\begin{eqnarray}
\xi &=& \frac{M'_x+iM'_y}{M+M'_z},\label{defximp}
\end{eqnarray}
but $M'_i$ are now real.
The question is: what is the equation satisfied by $\bfM'$?
Let us calculate the derivative of  $\bfM'$ with respect to $t$.
We have
\begin{eqnarray*}
\dot{M}'_x &=& 
\frac{\dot{\xi}-\dot{\xi}(\xi^*)^2+\dot{\xi}^*-
   \dot{\xi}^*\xi^2}{(|\xi|^2+1)^2} M,\\
\dot{M}'_y &=& -i 
\frac{\dot{\xi}+\dot{\xi}(\xi^*)^2-\dot{\xi}^*-
   \dot{\xi}^*\xi^2}{(|\xi|^2+1)^2} M,\\
\dot{M}'_z &=& -2 \frac{\dot{\xi}\xi^*+\xi\dot{\xi}^*}{(|\xi|^2+1)^2} M.
\end{eqnarray*}
If we express $\dot{\xi}$ and $\dot{\xi}^*$ through
equation (\ref{dotxiN2}) and its conjugate, we obtain
$\dotbfM'$ in terms of $\xi$ and $\xi^*$. If we
replace them by equation (\ref{defximp}) and its conjugate,
we obtain, after a lengthy but straightforward calculation,
\begin{eqnarray*}
\dot{M}'_x &=& -\frac{\gamma}{1+\alpha^2}(B_z M'_y- B_y M'_z)
-\frac{\alpha\gamma}{(1+\alpha^2)M}
\\&&\times
  (B_y M'_x M'_y - B_x {M'_y}^2 + B_z M'_x M'_z - B_x {M'_z}^2)
\\
\dot{M}'_y &=& -\frac{\gamma}{1+\alpha^2}(B_x M'_z- B_z M'_x)
-\frac{\alpha\gamma}{(1+\alpha^2)M}
\\&&\times
  (B_z M'_y M'_z - B_y {M'_z}^2 + B_x M'_y M'_x - B_y {M'_x}^2)
\\
\dot{M}'_z &=& -\frac{\gamma}{1+\alpha^2}(B_y M'_x- B_x M'_y)
-\frac{\alpha\gamma}{(1+\alpha^2)M}
\\&&\times
  (B_x M'_z M'_x - B_z {M'_x}^2 + B_y M'_z M'_y - B_z {M'_y}^2).
\end{eqnarray*}
This can be rewritten
\begin{eqnarray*}
\dotbfM' &=& - \frac{\gamma}{1+\alpha^2} \bfM'\times\bfB - 
   \frac{\gamma\alpha}{(1+\alpha^2)M} 
 \bfM'\times(\bfM'\times\bfB).
\end{eqnarray*}
We recognize the Landau-Lifshitz equation, in a form equivalent to the
LLG equation. To show the equivalence, use the LLG equation to derive
$\bfM'\times\dotbfM'=-\gamma  \bfM'\times(\bfM'\times\bfB) - \alpha M
\dotbfM'$,
introduce this expression in the LLG equation and solve for $\dotbfM'$.
Thus, we have transformed a solution $\bfM$ of the Bloch equation
into a solution $\bfM'$ of the LLG equation \eqref{LLGeq}.

We discuss now the simplest example of this transformation, to show 
how the method works in practice. We consider a constant external
magnetic field along the $z$ axis, i.e. $\bfB(t)=(0,0,B_0)$.
The solution of the Bloch equation is:
\begin{eqnarray*}
M_x &=& M \sin\theta_0 \cos(\Omega t + \phi_0),\\
M_y &=& M \sin\theta_0 \sin(\Omega t + \phi_0),\\
M_z &=& M \cos\theta_0,
\end{eqnarray*}
where $M$, $\theta_0$ and $\phi_0$ are constants and
$\Omega=\gamma B_0$.
$\bfM$ is obviously an analytic function of $\gamma$
and we can substitute $\gamma/(1-i\alpha)$ for
$\gamma$ in $\bfM$ to define the vector $\bfN$.
This gives us
\begin{eqnarray*}
\xi &=& \tan\frac{\theta_0}{2} \ee^{i(\Omega' t+\phi_0)}
            \ee^{-\alpha\Omega' t},
\end{eqnarray*}
with $\Omega'=\Omega/(1+\alpha^2)$ and we
recover the well-known solution of the LLG equation
\begin{eqnarray*}
M'_x &=& M \frac{\sin\theta_0\cos(\Omega' t+\phi_0)}{
         \cosh(\alpha\Omega' t)+ \cos\theta_0\sinh(\alpha\Omega' t)},\\
M'_y &=& M \frac{\sin\theta_0\sin(\Omega' t+\phi_0)}{
         \cosh(\alpha\Omega' t)+ \cos\theta_0\sinh(\alpha\Omega' t)},\\
M'_z &=& M \frac{\cos\theta_0\cosh(\alpha\Omega' t)+
                       \sinh(\alpha\Omega' t)}{
         \cosh(\alpha\Omega' t)+ \cos\theta_0\sinh(\alpha\Omega' t)}.
\end{eqnarray*}
If $\Omega>0$, the equilibrium magnetization is 
$\bfM'=(0,0,M)$; if $\Omega <0$, it is $\bfM'=(0,0,-M)$.
As expected, damping transforms a precession dynamics into
a motion towards an equilibrium state.

It is convenient to determine 
directly the influence of damping on the
dynamics of the two-level system described in the
Schr\"odinger or Heisenberg picture.
For a two-level system, the Schr\"odinger equation
is
\begin{eqnarray*}
i\hbar \frac{\dd\psi}{\dd t} &=& H(t) \psi,
\end{eqnarray*}
where $\psi$ has two components $\psi_1$ and $\psi_2$.
The Hamiltonian can be written
$H(t)=(\hbar\gamma/2)(B_0(t)+\sum_j B_j(t)\sigma_j)$, where
the constant $\gamma$ has been added for later convenience
and where $\sigma_j$ are the Pauli matrices, so that
$\sigma_a\sigma_b=\delta_{ab}+i\epsilon_{abc}\sigma_c$.
Defining $f(t)=\exp(-i\gamma\int_0^t B_0(\tau)\dd\tau)$
and $\psi(t)=f(t)\psi'(t)$ turn the Schr\"odinger equation for
$\psi$ into a Schr\"odinger equation for $\psi'$ with
the Hamiltonian $H=(\hbar\gamma/2)\sum_j B_j\sigma_j$.
Thus, without loss of generality, we use the latter
Hamiltonian.

Following Feynman and coll. \cite{FeynmanJAP}, 
the relation between the
density matrix $\rho$ (with matrix elements
$\rho_{ij}=\psi_i\psi_j^*$) and the
magnetic moment $\bfM$ is $M_x=\rho_{12}+\rho_{21}$,
$M_y=i(\rho_{12}-\rho_{21})$ and $M_z=\rho_{11}-\rho_{22}$.
Thus, we obtain
$\rho=(1/2)(1+\sum_j M_j\sigma_j)$ and $|\bfM|=1$.
We see that $\xi=(M_x+iM_y)/(1+M_z)=\psi_2/\psi_1$.
If we diagonalize $\rho$ we recover the states
$\psi_1$ and $\psi_2$ up to a phase that cannot be specified easily.
Therefore, we shall work with the density matrix.
The equation of motion for $\rho$ is
\begin{eqnarray*}
\frac{\dd \rho}{\dd t} &=& -\frac{i}{\hbar} {[}H,\rho{]}.
\end{eqnarray*}
The commutation relations for the Pauli matrices turn this equation
into the Bloch equation
\begin{eqnarray*}
\dotbfM &=& -\gamma \bfM\times\bfB.
\end{eqnarray*}

To turn the Bloch equation into the LLG equation, we just replace
$\bfM$ by $\bfM'$ and
$\bfB$ by $\bfB-(\alpha/\gamma) \dotbfM'$. If we denote by
$\rho'$ the density matrix corresponding to $\bfM'$,
we find the equation of motion in the presence of damping
\begin{eqnarray*}
\dotrho' &=& -\frac{i}{\hbar} {[}H,\rho'{]}+
   i \alpha {[}\dotrho',\rho'{]}.
\end{eqnarray*}
In other words, the damping term of the LLG equation
is transformed into a nonlinear term $i \alpha {[}\dotrho',\rho'{]}$
in the equation of motion of the density matrix.

If we replace $\bfB$ by $\bfB-(\alpha/\gamma) \dotbfM'$ in the Schr\"odinger
equation itself, we obtain the nonlinear Schr\"odinger equation
\begin{eqnarray*}
\frac{\dd\psi'_1}{\dd t} &=& 
\big((1+2i\alpha|\psi'_2|^2)B_3-i\alpha \psi_1(\psi'_2)^* B_+\big)\psi'_1
\\&&\vspace*{10mm}
+(1+i\alpha|\psi'_2|^2)B_- \psi'_2,\\
\frac{\dd\psi'_2}{\dd t} &=& 
-\big((1+2i\alpha|\psi'_1|^2)B_3+i\alpha (\psi'_1)^*\psi'_2 B_-\big)\psi'_2
\\&&\vspace*{10mm}
+(1+i\alpha|\psi'_1|^2)B_+ \psi'_1.
\end{eqnarray*}
Therefore, our method transforms the analytic solution of a Bloch
equation into the analytic solution of a nonlinear Schr\"odinger
equation.
We also reach the surprising conclusion that the nonlinear terms
in the nonlinear Schr\"odinger equation can describe the influence
of damping. However, this damping does not create
decoherence: it transforms $\bfM$ into $\bfM'$, which is real and 
satisfies $|\bfM'|=1$. As a consequence,
${(\rho')}^2 = \rho'$ and $\rho'$ is the 
density matrix of a pure state.

We investigate now the effect of damping
on two famous nonperturbative effects in
two-level systems: self-induced transparency
and dynamical localization.

McCall and Hahn \cite{McCall} discovered a solution
for the hyperbolic secant pulse $\bfB=(a/\cosh(t/\tau),0,0)$.
We recall that $\int_{-\infty}^\infty \dd t/\cosh(t/\tau)=\tau\pi$.
We consider more generally a spin system
submitted to a time-varying magnetic field linearly polarized
along $Ox$:
$\bfB=(b(t),0,0)$, where $b(t)$ is only required to be integrable.
At time $t=0$ the spin has the spherical coordinates $\theta_0$ and
$\phi_0$.  Let $f(t)=\int_{t_0}^t \dd\tau b(\tau)$
and $a=(1-\xi_0)/(1+\xi_0)$ with $\xi_0=\tan(\theta_0/2) \ee^{i\phi_0}$.
The solution of the Bloch equation is
\begin{eqnarray*}
M_x(t) &=& M \frac{1 - \rho^2}{1 + \rho^2},\\
M_y(t) &=& -2 M \frac{\rho\sin(x(t)+u)}{1 + \rho^2},\\
M_z(t) &=& 2 M \frac{\rho\cos(x(t)+u)}{1 + \rho^2},
\end{eqnarray*}
with $\rho=|a|$, $a=\rho \ee^{iu}$ and $x(t)=\gamma f(t)$.
Self-induced transparency occurs when
$x(\infty)=2n\pi$ because, after a long interaction
with the external field, the system finds itself in its state
at $t_0$.
In the case of the McCall and Hahn pulse, we find
$f(t)=2 a \tau \atan\big(\tanh((t-t_0)/2\tau)\big)$, which tends
to $a\tau\pi/2$ for large $t$.
Therefore, self-induced transparency occurs
when $\gamma a \tau=4n$ for some integer $n$.

To determine the effect of damping on this phenomenon, we
calculate
\begin{eqnarray*}
\xi &=& \frac{1-a \ee^{i\bargamma f(t)}}{1+a \ee^{i\bargamma f(t)}},
\end{eqnarray*}
and the corresponding solution of the LLG equation is
\begin{eqnarray*}
M'_x(t) &=& M \frac{\ee^{2\alpha \barx(t)} - \rho^2}{\ee^{2\alpha \barx(t)}
 + \rho^2},\\
M'_y(t) &=& -2 M \frac{\ee^{\alpha \barx(t)}\rho\sin(\barx(t)+u)}
    {\ee^{2\alpha \barx(t)} + \rho^2},\\
M'_z(t) &=& 2 M \frac{\ee^{\alpha \barx(t)}\rho\cos(\barx(t)+u)}
    {\ee^{2\alpha \barx(t)} + \rho^2},
\end{eqnarray*}
with $\barx(t)=\gamma f(t)/(1+\alpha^2)$.
Two effects can be observed. Firstly, the resonance condition
is shifted from $x(\infty)=2n\pi$ to $x(\infty)=2n\pi(1+\alpha^2)$;
secondly, even at resonance the initial state
is not fully recovered.
For example, if the system is initially in the state
$\psi_1(0)=1$, $\psi_2(0)=0$, we have 
$M=1$, $\rho=1$ and $u=0$ and the final state is,
at resonance,
\begin{eqnarray*}
M'_x(\infty) &=& \tanh(\alpha\barx(\infty)),\\
M'_y(\infty) &=& 0,\\
M'_z(\infty) &=& \frac{1}{\cosh(\alpha\barx(\infty))}.
\end{eqnarray*}
Thus, it is not possible to recover the initial state
because $\alpha \barx(\infty)\not=0$.
Therefore, although the damping of the LLG equation
amounts only to a trend towards an equilibrium state and 
not to a decoherence, it leads to a loss
of self-induced transparency.
However, if $\alpha$ is small, this loss is reasonably small
and the reader might have the feeling that damping
has only a minor effect on the dynamics.
The next example shows that this is not the case.

Quantum systems under the influence of a periodic or
quasiperiodic external field can exhibit a
freezing of the diffusion of its quantum state
\cite{Grifoni}, called \emph{dynamical localization}. 
This effect is explained by the Floquet structure of 
its quantum state.  
It was analyzed rigorously for a two-level system
with $\bfB=(-f(t),0,\epsilon)$ by Barata and Cortez
\cite{BarataJMP}.
In the case where $f(t)=a \cos\omega t$, they
showed that
the crucial parameter is $\langle q^2\rangle$, the time average
of $q^2(t)$, where
$q(t) = \exp\big(i a \gamma\sin(\omega t)/\omega\big)$.
Dynamical localization occurs when
$\langle q^2\rangle=J_0(2a \gamma/\omega)$  is zero, i.e. when
$\chi=2a\gamma/\omega$ is a zero of the Bessel function $J_0$.
In that case, dynamic localization
is described as a modulation with
a very slow secular frequency 
$\Omega = 2 \gamma^3\epsilon^3 T(\chi)/\omega^2 + O(\epsilon^4)$,
with $T(\chi) = -\sum_{m,n} J_n(\chi)J_{n-m}(\chi)J_m(\chi)/(mn)$,
where $J_n$ are Bessel functions and the sum is over all
nonzero integers.

In the presence of damping, $\gamma$ is replaced by 
$\gamma/(1-i\alpha)$ and $\langle q^2\rangle$ can no longer be
zero because all zeros of $J_0$ are real \cite{Watson}.
Thus, contrary to the case of self-induced transparency,
no resonance is strictly possible in the presence of damping.
However, a second more dramatic effect occurs.  If we take
$\chi=2a\gamma/\omega$ a zero of the Bessel function $J_0$,
and expand $\langle q^2\rangle$ around $\lambda=\chi/(1+\alpha^2)$,
the Bessel expansion\cite{Abramowitz} yields
\begin{eqnarray*}
\langle q^2\rangle &=& J_0\big(\lambda(1+i\alpha)\big)
\\&=&
 J_0(\lambda) J_0(i\lambda\alpha)+2 \sum_{k=1}^\infty (-1)^k
   J_{k}(\lambda) J_k(i\lambda\alpha).
\end{eqnarray*}
Therefore,
\begin{eqnarray*}
\langle q^2\rangle &=&  -i\chi J_1(\chi)\alpha 
-(\chi/2)^2J_2(\chi)\alpha^2
+ O(\alpha^3).
\end{eqnarray*}
The off-resonance secular frequency is \cite{BarataJMP} 
$\Omega=\epsilon \gamma\langle q^2\rangle$, so that
\begin{eqnarray*}
\Omega &=& 
 -i\epsilon\gamma\alpha \chi J_1(\chi) 
  - \epsilon \gamma\alpha^2 (\chi/2)^2J_2(\chi) + O(\alpha^3).
\end{eqnarray*}
The factor $J_1(\chi)$ is not zero because
the Bessel functions $J_0$ and $J_1$ have no
common zeros \cite{Watson}.
In other words, the secular frequency acquires
an imaginary part. Barata analyzed the
influence of an imaginary part in detail \cite{BarataRMP}
and showed that the Floquet structure of the wavefunction
generates exponentially increasing terms that
destroy the unitarity of the quantum evolution.
We have shown that the norm of the wavefunction is
conserved by the damped dynamics. Thus, the quantum
evolution is still unitary and damping destroys
the Floquet structure of the wavefunction. In this 
last example, damping has a dramatic effect on
the dynamics of the system.
Indeed, experiments confirm that dynamical
localization is very sensitive to external
perturbations \cite{Lignier}.

It was shown in this paper that analytic solutions
of the Bloch equations can be transformed into
analytic solutions of the LLG equation. The 
influence of damping leads
to interesting results for two-level systems,
in particular the transformation of the 
linear into the non linear Schr\"odinger equation.
The effect of damping was shown to be
quite different for self-induced transparency,
where it is moderate, and for dynamical
localization, where it destroys
the structure of the dynamics.

\begin{acknowledgments}
I am very grateful to J. Goulon for discussions
on the LLG equation.
\end{acknowledgments}

\end{document}